\documentclass[sigchi,screen]{acmart}

\AtBeginDocument{%
  \providecommand\BibTeX{{%
    \normalfont B\kern-0.5em{\scshape i\kern-0.25em b}\kern-0.8em\TeX}}}

\setcopyright{rightsretained}
\acmPrice{15.00}
\acmDOI{00.0000/0000000.0000000}
\acmYear{2025}
\copyrightyear{2025}
\acmSubmissionID{tbd} 
\acmISBN{000-0-0000-0000-0/00/00}
\acmConference[CHI '25 Workshop: Everyday AR through AI-in-the-Loop]{Workshop Proceedings of the 2025 Everyday AR through AI-in-the-Loop Workshop}{April 2025}{Yokohama, Japan}
\acmBooktitle{tbd}

 \usepackage{microtype} 

\begin{document}

\def\name{A Vision for AI-Driven Adaptation of Dynamic AR Content to Users and Environments}
\title[\name]{\name{} }


\author{Julian Rasch}
\email{julian.rasch@ifi.lmu.de}
\affiliation{%
  \institution{LMU Munich}
  \city{Munich}
  \country{Germany}
}

\author{Florian Müller}
\email{florian.mueller@tu-darmstadt.de}
\affiliation{%
  \institution{TU Darmstadt}
  \city{Darmstadt}
  \country{Germany}
}

\author{Francesco Chiossi}
\email{francesco.chiossi@ifi.lmu.de}
\affiliation{%
  \institution{LMU Munich}
  \city{Munich}
  \country{Germany}
}

\renewcommand{\shortauthors}{Rasch et al.}

\begin{abstract}
Augmented Reality (AR) is transforming the way we interact with virtual information in the physical world. By overlaying digital content in real-world environments, AR enables new forms of immersive and engaging experiences. However, existing AR systems often struggle to effectively manage the many interactive possibilities that AR presents.
%
This vision paper speculates on AI-driven approaches for adaptive AR content placement, dynamically adjusting to user movement and environmental changes.
By leveraging machine learning methods, such a system would intelligently manage content distribution between AR projections integrated into the external environment and fixed static content, enabling seamless UI layout and potentially reducing users' cognitive load.
By exploring the possibilities of AI-driven dynamic AR content placement, we aim to envision new opportunities for innovation and improvement in various industries, from urban navigation and workplace productivity to immersive learning and beyond. This paper outlines a vision for the development of more intuitive, engaging, and effective AI-powered AR experiences.
\end{abstract}

\ccsdesc[500]{Human-centered computing~Ubiquitous and mobile devices}
\ccsdesc[300]{Human-centered computing~Interaction techniques}
\ccsdesc[300]{Human-centered computing~Interactive systems and tools}

\keywords{Artificial Intelligence, Mixed Reality, Augmented Reality, Locomotion, Intelligent User Interfaces}



\maketitle

\section{Motivation}

The integration of AI capabilities into AR technologies presents new opportunities to solve a key challenge in spatial computing, namely adapting AR content to dynamic users and environments as they move through space. Traditional AR systems often use static content placement or fixed positioning within users' field-of-view, limiting their potential in dynamic, real-world scenarios \cite{lages_walking_2019}.
Recent research has explored various approaches to this challenge. Tanaka et al. \cite{kohei_tanaka_information_2008} proposed a layout method for AR displays in changing environments. More recent systems like SituationAdapt \cite{li_situationadapt_2024} use large language models (LLMs) to optimize mixed reality interfaces based on context.
Further, UI positioning in AR has seen advancements, with studies exploring user preferences for video call windows in different environments \,\cite{chang_exploring_2024}, user performance in dual-task walking scenarios with varying AR content anchoring around the user's body\,\cite{rasch_AR_you_2025}, and techniques for UI transitions between static, dynamic, and "self entities" \cite{pei_ui_2024}. Research has also identified areas where AI enhances AR, including pattern detection and improved experiences through neural networks and natural language processing \cite{reiners_combination_2021, hirzle_when_2023}.


While these examples show progress on individual aspects of AR interfaces, it is often limited to these specific scenarios. For AR to become our primary interaction medium, its interfaces must be suitable for a wide range of everyday situations, including indoors and outdoors, while walking or standing, when alone, or in crowded spaces. To bridge the gap between these localized solutions and create more generalizable, adaptable interfaces that can seamlessly integrate with dynamic environments and users, our work proposes an AI-driven system for dynamic AR content placement that adapts to these changes 
This approach manages content distribution between environmental projections and head-mounted displays (HMDs), offering more nuanced and context-aware placement.

\section{Problem Space}
\begin{figure}[t]
    \centering
    \includegraphics[width=1.0\linewidth]{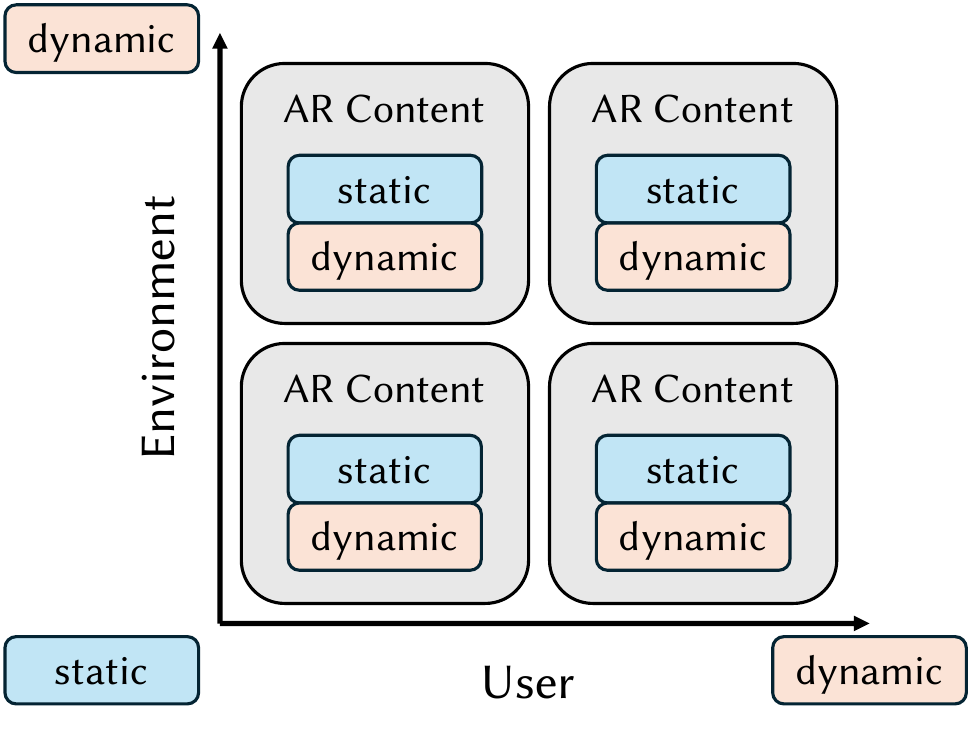}
    \caption{The design space for everyday AR interactions categorizes them into use cases based on user mobility and environmental stability each with dynamic and static AR content.}
    \label{fig:Content-User-Environement}
    \Description{
    }
\end{figure}

When studying and designing AR interactions for users in everyday scenarios, we can separate use cases based on static or dynamic characteristics of users and environment into a matrix with four distinct scenarios as shown in \autoref{fig:Content-User-Environement}. In each given scenario, the AR content, can be static, dynamic, or also combine AR elements of both characteristics. Adaptation of UI elements might be necessary because of visual changes in the user's field of view. These changes occur in particular as a result of spatial changes — either by the users themselves (through head rotation, locomotion, etc.) or by elements of the environment (cars, people, etc.). 




\textbf{Static User, Static Environment} Both user and environment are fixed, e.g., in scenarios like inspecting static AR models. Static content supports applications where consistent information delivery is important, such as text labels on artifacts or fixed guidance displays in training. Dynamic content, if used, could introduce adaptive layers for exploration but may not always be necessary when precision and stability are prioritized.

\textbf{Static User, Dynamic Environment} A stationary user interacts in a changing environment, such as viewing AR content in public transport. Static content is beneficial for persistent notifications or content providing general information, such as route maps or advertisements. However, dynamic content could enhance situational relevance by responding to environmental changes like crowd density or lighting conditions.

\textbf{Dynamic User, Static Environment} A moving user in a stable environment interacts with fixed AR content, such as navigation waypoints in museums or airports. Static content, like directional arrows or labels, remains effective for guiding users without overwhelming them. The challenge lies in maintaining visibility across varying perspectives. Dynamic content can offer adaptive interfaces, such as repositioning navigation cues or dynamically adjusting information displays based on user location or movement patterns.

\textbf{Dynamic User, Dynamic Environment} Both user and environment are in motion, such as AR shopping in busy spaces. Static content, such as price tags or simple notifications, remains useful for persistent information. However, its effectiveness may decline as the environment becomes more chaotic. Applications like interactive shopping assistants or context-based offers highlight how dynamic content can personalize the user experience without relying solely on environmental adaptation.

Each scenario in \autoref{fig:Content-User-Environement} comes with unique challenges. However, out of these combinations, a dynamic user in a dynamic environment with dynamic AR content is the least researched and most challenging field.
The central question here is: How can we design an AR system that can seamlessly integrate with the dynamic environment and provide a coherent and engaging user experience? Specifically, what content should be displayed as part of the augmented environment, where and how should it be overlaid, and what should remain static within the user’s field of view?

\section{Vision for an AI-powered AR System for Dynamic AR Content for Users in Dynamic Environments} 


For this most challenging case of dynamic users in dynamic environments with dynamic AR content, multiple solutions approaches exist that could be integrated into an AI-powered AR system, which we envision as follows.

The system can create \textbf{contextual content} for users utilizing computer vision for scene understanding and LLMs for the generation of contextual information and 3D content\,\cite{de2024llmr}.
Traditional AR content placement often focuses on either the user's perspective or static environmental elements. In contrast, the AI-powered AR system has \textbf{environmental awareness} and considers both the user's movement and changes in the surrounding environment, creating a more holistic and responsive system.
Unlike traditional methods that often rely on static content placement or fixed positioning within the user's field of view, the envisioned system could \textbf{dynamically adapt} content placement based on real-time user movement and environmental changes.


The system employs a multi-modal approach to \textbf{manage content} distribution between environmental projections and HMD visualizations intelligently. Reinforcement learning algorithms optimize \textbf{content placement} by considering user engagement and task performance metrics, while computer vision models identify suitable projection surfaces in the environment. Natural language processing techniques help prioritize content relevance, ensuring that information is presented in the most appropriate format and location. This adaptive approach is expected to significantly reduce cognitive load by \textbf{distributing content} between environmental "white spaces" and personal HMDs. For instance, time-critical information might appear on the HMD, while contextual details are projected onto nearby surfaces. Unlike traditional single-display methods that often overwhelm users with dense information, this dynamic distribution strategy \textbf{minimizes visual clutter} while maintaining information accessible. The result is a more intuitive and less cognitively demanding AR experience that adapts to both user needs and environmental constraints.


The AI-powered AR system incorporates a deeper \textbf{understanding of context}, potentially considering factors such as user intent, environmental constraints, and task relevance when placing content. Traditional methods often lack this level of contextual awareness.
The AI-powered AR system concept aims to create a more \textbf{seamless integration} of digital content into physical spaces where possible, potentially transforming how users interact with information in their daily lives. This goes beyond the often compartmentalized approach of traditional AR content placement.

While some of these components already exist individually, and others are the subject of current research, there is no such combined system yet. This envisioned AI-powered approach for content placement in dynamic AR environments could be a significant advancement in creating more intuitive, adaptive, and user-centric AR experiences for most everyday AR use-cases.


\bibliographystyle{ACM-Reference-Format}
\bibliography{references2}


\begin{thebibliography}{9}


\ifx \showCODEN    \undefined \def \showCODEN     #1{\unskip}     \fi
\ifx \showDOI      \undefined \def \showDOI       #1{#1}\fi
\ifx \showISBNx    \undefined \def \showISBNx     #1{\unskip}     \fi
\ifx \showISBNxiii \undefined \def \showISBNxiii  #1{\unskip}     \fi
\ifx \showISSN     \undefined \def \showISSN      #1{\unskip}     \fi
\ifx \showLCCN     \undefined \def \showLCCN      #1{\unskip}     \fi
\ifx \shownote     \undefined \def \shownote      #1{#1}          \fi
\ifx \showarticletitle \undefined \def \showarticletitle #1{#1}   \fi
\ifx \showURL      \undefined \def \showURL       {\relax}        \fi
\providecommand\bibfield[2]{#2}
\providecommand\bibinfo[2]{#2}
\providecommand\natexlab[1]{#1}
\providecommand\showeprint[2][]{arXiv:#2}

\bibitem[Chang et~al\mbox{.}(2024)]%
        {chang_exploring_2024}
\bibfield{author}{\bibinfo{person}{Chiao-Ju Chang}, \bibinfo{person}{Yu~Lun Hsu}, \bibinfo{person}{Wei Tian~Mireille Tan}, \bibinfo{person}{Yu-Cheng Chang}, \bibinfo{person}{Pin~Chun Lu}, \bibinfo{person}{Yu Chen}, \bibinfo{person}{Yi-Han Wang}, {and} \bibinfo{person}{Mike~Y. Chen}.} \bibinfo{year}{2024}\natexlab{}.
\newblock \showarticletitle{Exploring {Augmented} {Reality} {Interface} {Designs} for {Virtual} {Meetings} in {Real}-world {Walking} {Contexts}}. In \bibinfo{booktitle}{\emph{Designing {Interactive} {Systems} {Conference}}}. \bibinfo{publisher}{ACM}, \bibinfo{address}{IT University of Copenhagen Denmark}, \bibinfo{pages}{391--408}.
\newblock
\showISBNx{9798400705830}
\urldef\tempurl%
\url{https://doi.org/10.1145/3643834.3661538}
\showDOI{\tempurl}


\bibitem[De~La~Torre et~al\mbox{.}(2024)]%
        {de2024llmr}
\bibfield{author}{\bibinfo{person}{Fernanda De~La~Torre}, \bibinfo{person}{Cathy~Mengying Fang}, \bibinfo{person}{Han Huang}, \bibinfo{person}{Andrzej Banburski-Fahey}, \bibinfo{person}{Judith Amores~Fernandez}, {and} \bibinfo{person}{Jaron Lanier}.} \bibinfo{year}{2024}\natexlab{}.
\newblock \showarticletitle{Llmr: Real-time prompting of interactive worlds using large language models}. In \bibinfo{booktitle}{\emph{Proceedings of the 2024 CHI Conference on Human Factors in Computing Systems}}. \bibinfo{pages}{1--22}.
\newblock


\bibitem[Hirzle et~al\mbox{.}(2023)]%
        {hirzle_when_2023}
\bibfield{author}{\bibinfo{person}{Teresa Hirzle}, \bibinfo{person}{Florian Müller}, \bibinfo{person}{Fiona Draxler}, \bibinfo{person}{Martin Schmitz}, \bibinfo{person}{Pascal Knierim}, {and} \bibinfo{person}{Kasper Hornbæk}.} \bibinfo{year}{2023}\natexlab{}.
\newblock \showarticletitle{When {XR} and {AI} {Meet} - {A} {Scoping} {Review} on {Extended} {Reality} and {Artificial} {Intelligence}}. In \bibinfo{booktitle}{\emph{Proceedings of the 2023 {CHI} {Conference} on {Human} {Factors} in {Computing} {Systems}}}. \bibinfo{publisher}{ACM}, \bibinfo{address}{Hamburg Germany}, \bibinfo{pages}{1--45}.
\newblock
\showISBNx{978-1-4503-9421-5}
\urldef\tempurl%
\url{https://doi.org/10.1145/3544548.3581072}
\showDOI{\tempurl}


\bibitem[{Kohei Tanaka} et~al\mbox{.}(2008)]%
        {kohei_tanaka_information_2008}
\bibfield{author}{\bibinfo{person}{{Kohei Tanaka}}, \bibinfo{person}{{Yasue Kishino}}, \bibinfo{person}{{Masakazu Miyamae}}, \bibinfo{person}{{Tsutomu Terada}}, {and} \bibinfo{person}{{Shojiro Nishio}}.} \bibinfo{year}{2008}\natexlab{}.
\newblock \showarticletitle{An information layout method for an optical see-through head mounted display focusing on the viewability}. In \bibinfo{booktitle}{\emph{2008 7th {IEEE}/{ACM} {International} {Symposium} on {Mixed} and {Augmented} {Reality}}}. \bibinfo{publisher}{IEEE}, \bibinfo{address}{Cambridge, UK}, \bibinfo{pages}{139--142}.
\newblock
\showISBNx{978-1-4244-2840-3}
\urldef\tempurl%
\url{https://doi.org/10.1109/ISMAR.2008.4637340}
\showDOI{\tempurl}


\bibitem[Lages and Bowman(2019)]%
        {lages_walking_2019}
\bibfield{author}{\bibinfo{person}{Wallace~S. Lages} {and} \bibinfo{person}{Doug~A. Bowman}.} \bibinfo{year}{2019}\natexlab{}.
\newblock \showarticletitle{Walking with adaptive augmented reality workspaces: design and usage patterns}. In \bibinfo{booktitle}{\emph{Proceedings of the 24th {International} {Conference} on {Intelligent} {User} {Interfaces}}}. \bibinfo{publisher}{ACM}, \bibinfo{address}{Marina del Ray California}, \bibinfo{pages}{356--366}.
\newblock
\showISBNx{978-1-4503-6272-6}
\urldef\tempurl%
\url{https://doi.org/10.1145/3301275.3302278}
\showDOI{\tempurl}


\bibitem[Li et~al\mbox{.}(2024)]%
        {li_situationadapt_2024}
\bibfield{author}{\bibinfo{person}{Zhipeng Li}, \bibinfo{person}{Christoph Gebhardt}, \bibinfo{person}{Yves Inglin}, \bibinfo{person}{Nicolas Steck}, \bibinfo{person}{Paul Streli}, {and} \bibinfo{person}{Christian Holz}.} \bibinfo{year}{2024}\natexlab{}.
\newblock \showarticletitle{{SituationAdapt}: {Contextual} {UI} {Optimization} in {Mixed} {Reality} with {Situation} {Awareness} via {LLM} {Reasoning}}. In \bibinfo{booktitle}{\emph{Proceedings of the 37th {Annual} {ACM} {Symposium} on {User} {Interface} {Software} and {Technology}}}. \bibinfo{publisher}{ACM}, \bibinfo{address}{Pittsburgh PA USA}, \bibinfo{pages}{1--13}.
\newblock
\showISBNx{979-8-4007-0628-8}
\urldef\tempurl%
\url{https://doi.org/10.1145/3654777.3676470}
\showDOI{\tempurl}


\bibitem[Pei et~al\mbox{.}(2024)]%
        {pei_ui_2024}
\bibfield{author}{\bibinfo{person}{Siyou Pei}, \bibinfo{person}{David Kim}, \bibinfo{person}{Alex Olwal}, \bibinfo{person}{Yang Zhang}, {and} \bibinfo{person}{Ruofei Du}.} \bibinfo{year}{2024}\natexlab{}.
\newblock \showarticletitle{{UI} {Mobility} {Control} in {XR}: {Switching} {UI} {Positionings} between {Static}, {Dynamic}, and {Self} {Entities}}. In \bibinfo{booktitle}{\emph{Proceedings of the {CHI} {Conference} on {Human} {Factors} in {Computing} {Systems}}}. \bibinfo{publisher}{ACM}, \bibinfo{address}{Honolulu HI USA}, \bibinfo{pages}{1--12}.
\newblock
\showISBNx{979-8-4007-0330-0}
\urldef\tempurl%
\url{https://doi.org/10.1145/3613904.3642220}
\showDOI{\tempurl}


\bibitem[Rasch et~al\mbox{.}(2025)]%
        {rasch_AR_you_2025}
\bibfield{author}{\bibinfo{person}{Julian Rasch}, \bibinfo{person}{Matthias Wilhalm}, \bibinfo{person}{Florian M\"uller}, {and} \bibinfo{person}{Francesco Chiossi}.} \bibinfo{year}{2025}\natexlab{}.
\newblock \showarticletitle{{AR} {You} on {Track}? {Investigating} {Effects} of {Augmented} {Reality} {Anchoring} on {Dual-Task} {Performance} {While} {Walking}}. In \bibinfo{booktitle}{\emph{Proceedings of the {CHI} {Conference} on {Human} {Factors} in {Computing} {Systems}}}. \bibinfo{publisher}{ACM}, \bibinfo{address}{Yokohama Japan}, \bibinfo{pages}{1--21}.
\newblock
\showISBNx{979-8-4007-1394-1}
\urldef\tempurl%
\url{https://doi.org/10.1145/3706598.3714258}
\showDOI{\tempurl}


\bibitem[Reiners et~al\mbox{.}(2021)]%
        {reiners_combination_2021}
\bibfield{author}{\bibinfo{person}{Dirk Reiners}, \bibinfo{person}{Mohammad~Reza Davahli}, \bibinfo{person}{Waldemar Karwowski}, {and} \bibinfo{person}{Carolina Cruz-Neira}.} \bibinfo{year}{2021}\natexlab{}.
\newblock \showarticletitle{The {Combination} of {Artificial} {Intelligence} and {Extended} {Reality}: {A} {Systematic} {Review}}.
\newblock \bibinfo{journal}{\emph{Frontiers in Virtual Reality}}  \bibinfo{volume}{2} (\bibinfo{date}{Sept.} \bibinfo{year}{2021}), \bibinfo{pages}{721933}.
\newblock
\showISSN{2673-4192}
\urldef\tempurl%
\url{https://doi.org/10.3389/frvir.2021.721933}
\showDOI{\tempurl}


\end{thebibliography}

\end{document}